# Human Cognition and Language Processing with Neural-Lexicon Hypothesis

Zang-Hee Cho[1†], Sun-Ha Paek [2], Young-Bo Kim [3], Taigyoun Cho[4], Hyejin Jeong[1], Haigun Lee [5†]

[1]*Neuroscience Convergence Center, Institute of Green Manufacturing Technology, Korea University, Seoul, South Korea*
[2]*Department of Neurosurgery, School of Medicine, Seoul National University, Seoul, South Korea*
[3]*Department of Neurosurgery, School of Medicine, Gachon University, Incheon, South Korea*
[4]*Department of Industrial Design, Hong Ik University, Seoul, South Korea*
[5]*Department of Material Science, Institute of Green Manufacturing Technology, Korea University, Seoul, South Korea*
[†]*These authors contributed equally to Co-Corresponding author*

**ABSTRACT**

Cognition and language seem closely related to the human cognitive process, although they have not been studied and investigated in detail. Our brain is too complex to fully comprehend the structures and connectivity, as well as its functions, with the currently available technology such as electro-encephalography, positron emission tomography, or functional magnetic resonance imaging, and neurobiological data. Therefore, the exploration of neurobiological processes, such as cognition, requires substantially more related evidences, especially from in-vivo human experiments. Cognition and language are of inter-disciplinary nature and additional methodological support is needed from other disciplines, such as deep learning in the field of artificial intelligence, for example. In this paper, we have attempted to explain the neural mechanisms underlying "cognition and language processing" or "cognition or thinking" using a novel neural network model with several newly emerging developments such as neuronal resonance, in-vivo human fiber tractography or connectivity data, Engram and Hebbian hypothesis, human memory formation in the high brain areas, deep learning, and more recently developed neural memory concepts, the neural lexicon. The neural lexicon is developed via language by repeated exposure to the neural system, similar to multilayer signal processing in deep learning. We have derived a neural model to explain how human "cognition and language processing" or "cognition and thinking" works, with a focus on language, a universal medium of the human society. Although the proposed hypothesis is not fully based on experimental evidences, a substantial portion of the observations in this study is directly and indirectly supported by recent experimental findings and the theoretical bases of deep learning research.

## 1. INTRODUCTION

Human cognition and language are the two most difficult subjects due to their wide variely and the lack of evidence especially on their neural basis. Nevertheless, it is worthwhile to study based on contenparary neuroscience perspective with the modern logic behind deep learning.

For this study, we have chosen "cognition" and "language," or slightly differently "cognition" and "thinking," as the central topics. We attempted to answer the questions from neuroscience perspectives while leveraging AI tools, specifically deep learning with multiple signal processing layers, which we modified with neural bases and termed as neural lexicons. Neural lexicons are memory units believed to be developed via languages and other related neuroscience developments such as newly developed fiber tractography or connectivity data neuronal Resonance, Hebbian neural processes, or the Engram[1–6]. Language, in particular, appears to play a major role and is a unique human property with which we have postulated and developed neural lexicons, which are similar to the signal processing layers in deep learning[7–15]. A general sketch of the proposed neural pathways or the network for "cognition and language" or "cognition and thinking" is provided in **Fig. 1**, where three neural Lexicons and two memory units are shown together with connecting fibers from the sensory areas to the motor areas. We have provided details of the neural circuit shown in **Fig. 1** subsequently.

We begin with the basic neuronal process as shown in **Fig. 2**. With this process, a conceptual neuronal activity involving previously excited and newly excited neurons, especially the neurons that are excited at the same time by the external stimulation, such as the Hebbian neurons, seem to play significant a role in the early stages of the neural signal processing of the sensory cortex.

As a simplified example, the neuronal signal, which we coined as the "sensogram" is assumed to represent the total number of activated neurons at a given moment, which is shown as (see also Fig. 2(a)),

*E-mail: zhcho36@gmail.com*

$\mathcal{A}(t) = f_{00} + f_{11} + f_{21} + f(t_{12}) + f(t_{345})$:
Total Activated Neurons .... {Eq.1} Sensogram

where $f_{00}, f_{11},$ and $f_{21}$ are the zero[th], first, and second order noises, respectively, while $f(t_{12})$ and $f(t_{345})$ are the signal components at the same time, respectively. From Eq. 1, the total number of concurrently activated neurons, the Hebbian neurons or Cluster, can be represented as neuronal resonance at the primary sensory cortex as follows,

$\mathcal{A}'(t) = f_{11} + f_{21} + f(t_{12}) + f(t_{345})$: Activated Neurons by Hebbian rule.... {Eq.2} Pre-Engram

**Figs. 2(a) and (b)** show the transition from the sensogram to pre-Engram and, finally, the Engram formation. From Eq. 2, the Hebbian neurons, from which one can derive the Engram via the Engram pattern $\overline{\mathcal{B}(t)}$ at the secondary sensory area, can be written as,

$\mathcal{B}(t) = f_{21} + f(t_{12}) + f(t_{345})$ .... {Eq.3} Engram

where $f(t_{12})$ and $f(t_{345})$ are the components of the phoneme while $f_{21}$ is the system noise.

Subsequently, the Engram produced is sent to the higher cortical areas, such as the inferior parietal cortex (IPC) or lobe. In these areas, it is assumed that some form of higher-order memory units, termed the neural Lexicons, are developed via language, as shown in Fig. 2(c). In the n-Lexicon, the incoming Engram signal is absorbed by neuronal resonance with a pattern stored in the n-Lexicon, which is developed via language. This results in the pre-Langram or phoneme, which consists of phonetic components as shown on the right of the figure[2,5,6]. The sum of phonetic components that form a phoneme or the pre-Langram is given as:

$\mathcal{C}(t) = f(t_{12}) + f(t_{345})$: Phoneme .... {Eq.4} Pre-Langram

where $f(t_{12})$ and $f(t_{345})$ are the essential phonetic components of the Phoneme. A phoneme is considered a phonologic element that forms the basis of a Morpheme or Word, the most diminutive linguistic form or the semantic atom in the linguistic field[7,9,10,12,13,15]. As will be shown, with these bases, we will describe how words, sentences, cognition, decisions, and, eventually, thinking can be formulated.

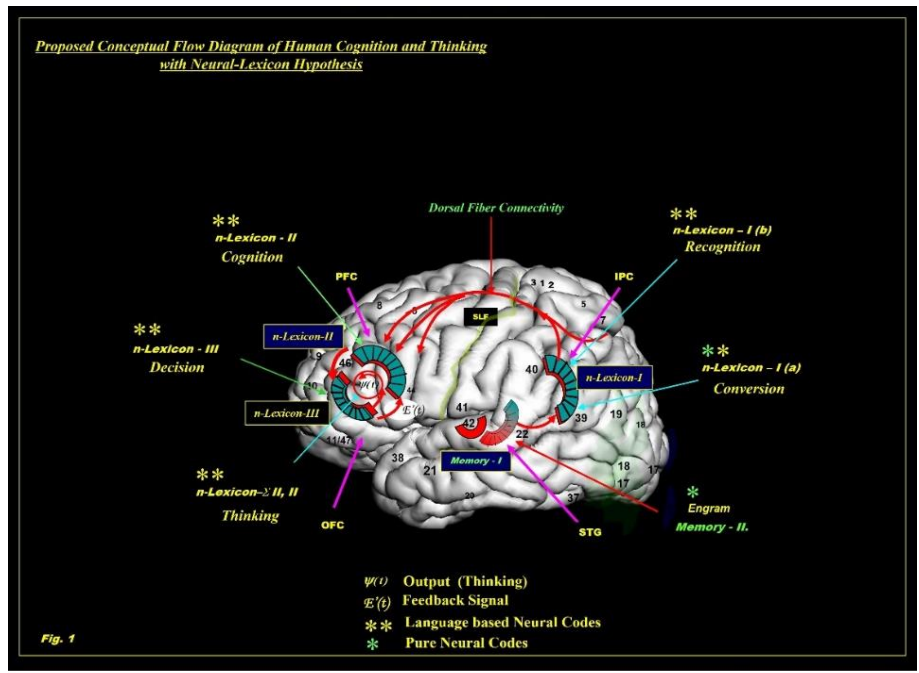

**Figure 1.** Proposed overall scheme of the human cognition and language processing with neural Lexicon hypothesis. Here we assumed three neural Lexicons with different functions, such as neural code conversion, grouping, sentence formation, and decision. Sentence formation leads to cognition and, therefore, "thinking," which is the highest cognitive function. * and ** represent the neural codes which comprises pure neuronal code and language based neural code, respectively. This model is based on recently obtained fiber tractography or connectivity, neuronal resonance, Engram and Hebbian neuronal cluster formation hypothesis, the neural Lexicon hypothesis or language-based memory formation we have proposed, and recently developed deep learning logic. These developments, although further experimental evidence is needed, allow us to hypothesize and conclude consistently and also explain how our human brain works during cognition and thinking, which are the highest cognitive functions.

Legends; STG: Superior Temporal Gyrus, IPC: Inferior Parietal Cortex, PFC: Prefrontal Cortex, OFC:

*E-mail: zhcho36@gmail.com

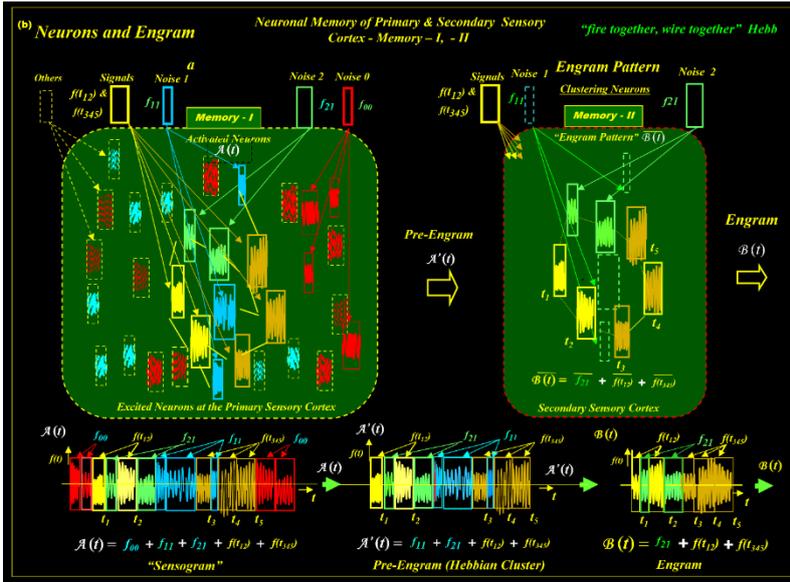
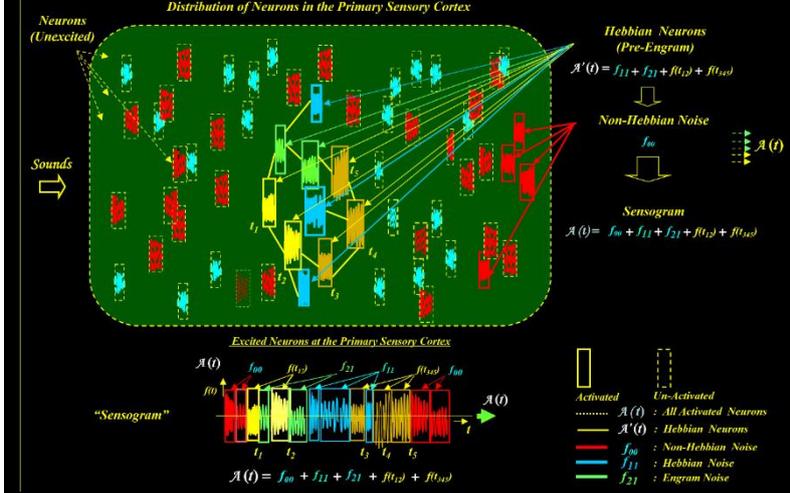
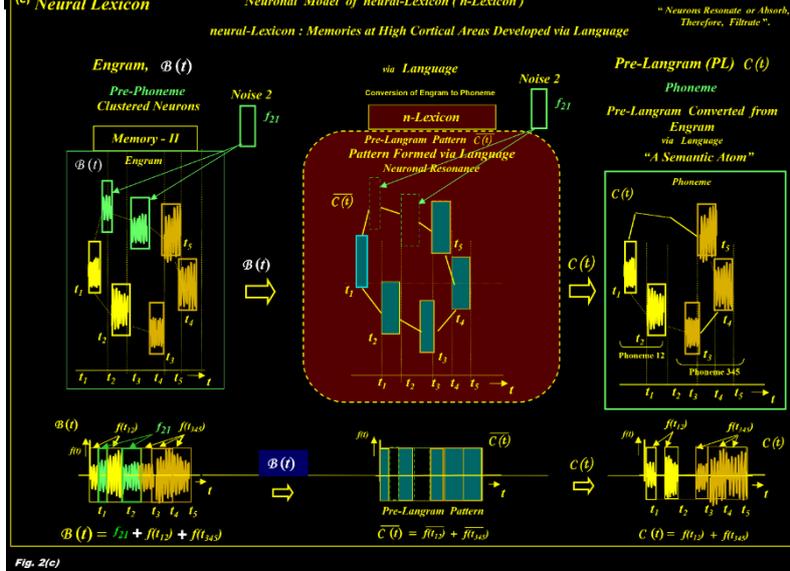

**Figure 2. (a).** A simplified neuronal model of activated and inactivated neurons that will generate neural codes such as the Engram is illustrated. From this, one can derive neuronal memories and neural Lexicons. Shown is a simple neuronal excitation model in the primary sensory area. Excited neurons represent noises and signals used for the cognition and thinking process. $f_{00}, f_{11},$ and $f_{21}$ are the noises while the $f(t_{12})$ and $f(t_{345})$ are the signal components. Note also that the Hebbian neuronal cluster, $\mathcal{A}'(t)$, will further be processed to form an Engram.

**(b).** Noisy input signal Sensogram, $\mathcal{A}(t)$, which includes the neuronal cluster formed by the Hebbian rule, is shown. When the input signal, $\mathcal{A}(t)$ or Sensorgram, including the Hebbian neuronal cluster ($\mathcal{A}'(t)$), is entered into Memory–II, where some form of a memory pattern, $\overline{\mathcal{B}(t)}$, is stored (ex. Engram pattern), incoming signals will be absorbed by neuronal resonance. This produces a new neuronal cluster called the Engram. Here we assumed $\mathcal{A}'(t)$ as a "pre-Engram," a neuronal cluster formed by the Hebbian rule.

**(c).** A certain memory pattern is formed and works as a filter for the incoming neuronal signals by neuronal resonance. Here the neural Lexicons assume a certain memory pattern which is formed by repeated language exposure. An exemplary n-Lexicon is shown, which involves a simple conversion of the incoming signal by neuronal resonance with help of a neural pattern, $\overline{\mathcal{C}(t)}$, formed via a language for example, pre-Langram for the conversion of Engram to phoneme. The functions of the n-Lexicons may vary with circumstances. n-Lexicons perform various major functions, from filtering or selection to organized grouping based on memory patterns developed via language.

Legends; $\mathcal{A}(t)$: Sensogram, total activated neurons including the Hebbian cluster and non-Hebbian neurons representing noises. $\mathcal{A}'(t)$: Hebbian Cluster. $\overline{\mathcal{B}(t)}$: A memory pattern of the Engram. $\mathcal{A}'(t)$: Hebbian Neuronal Cluster. $\mathcal{B}(t)$: Engram (Pre-Phoneme). $\overline{\mathcal{C}(t)}$: A memory pattern of a Phoneme (Pre-Langram). Langram is a temporary name assigned for "Language-Engram." $\mathcal{B}(t)$: Pre-Phoneme (Engram). $\mathcal{C}(t)$: Phoneme. $f(t_{12})$ and $f(t_{345})$: Components of a Phoneme. $f_{21}$: System Noise

*E-mail: zhcho36@gmail.com

## 2. MATERIALS AND MERHOD

**Neuronal Model for the Neuronal Code Processing - from Sensogram to Engram - Memory-I and II**

The overall model of the neural signal processing and the related human brain surface (left hemisphere or cortical surface) is shown in **Fig. 3-5**. Similar to the multi-layer deep learning process, we propose a neural process based on neuro-biological perspectives using newly proposed memory units, which are the neural Lexicons or n-Lexicons[14]. We have already discussed the basis in **Fig. 2(c)**, where we assumed that the input signal produces a neuronal cluster known as the Engram. The Engram was proposed as early as the beginning of the twentieth century by Semon, later by Hebb, and more recently by Tonegawa[2,5,6]. The Engram is conceived as the result of the input signal, which is the sensogram $\mathcal{A}(t)$; the sensogram includes the Hebbian cluster ($\mathcal{A}'(t)$) and other noises. When the filtered sensogram, which is the pre-Engram $\mathcal{A}'(t)$, enters the secondary sensory cortex, it is filtered by the memory ($\overline{\mathcal{B}(t)}$) an Engram pattern, resulting in the Engram ($\mathcal{B}(t)$), as shown in **Fig. 3(a)**.

**Numerical Approach from the Engram to Cognition via neural Lexicons, n-Lexicon-I and II**

After we have obtained the Engram, as shown in **Figs. 3(a) and (b),** we begin obtaining n-Lexicons with which those neural codes such as the Engram are converted to pseudo-language-based neural codes such as the phonemes or pre-Langrams. This first conversion process from the Engram to phoneme will take place at the front part of n-Lexicon-I (Ia), as shown in Fig. 3(b). This phoneme is composed of $f(t_{12})$ and $f(t_{345})$. In the following stage, n-Lexicon-I (Ib), a group of phonemes will take place. This will form a word or morpheme, which is a true language-based neural code; the "Word" with the neural pattern which is developed via language in n-Lexicon-I (Ib). The n-Lexicon-I, therefore, involves two signal processings, namely, the conversion of the Engram to the phoneme in n-Lexicon-I (Ia) and the subsequent formation of the "Word" by a grouping of the phonemes in n-Lexicon-I (Ib), which can be written as:

$\mathcal{C}'(t) = \mathcal{C}_1(t) + \mathcal{C}_2(t) + \mathcal{C}_3(t)$: Word (Group of Phonemes)…{Eq. 5} Langram or word

where $\mathcal{C}_1(t)$, $\mathcal{C}_2(t)$, and $\mathcal{C}_3(t)$ are the phonemes.

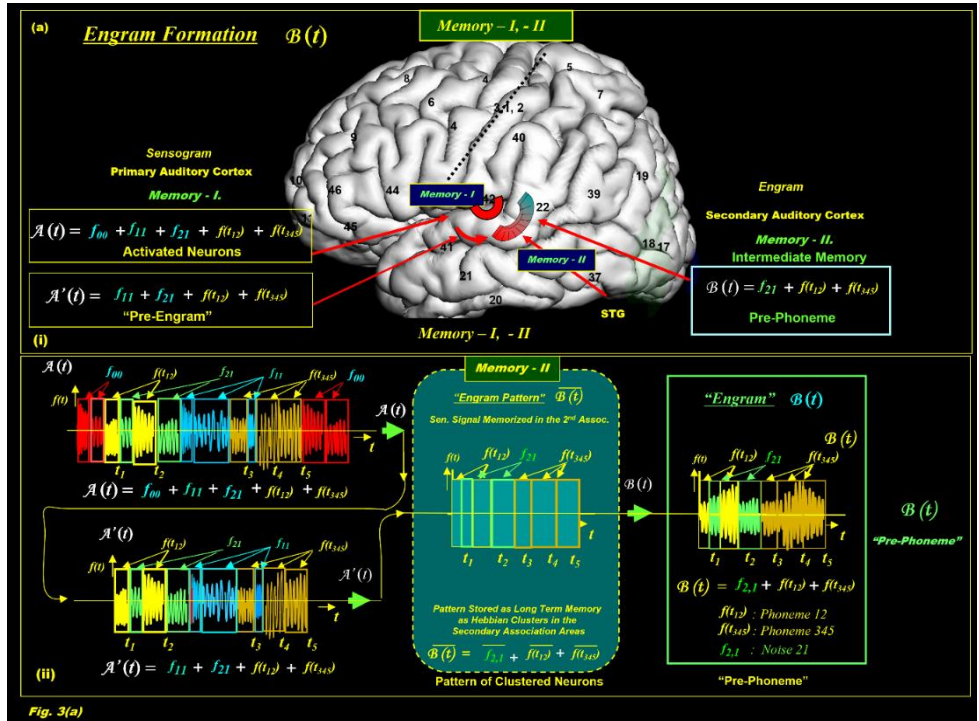

**Figure 3. (a).** The first parts of the sensory memory units, such as Memory-I and -II, are shown in the primary and secondary auditory areas. In the primary auditory area, excited neurons produce signals as shown in the bottom-left ($\mathcal{A}(t)$). Among the excited neurons, the Hebbian neurons ($\mathcal{A}'(t)$) transmit to the secondary sensory area, memory-II where memory is formed as an intermediate, long-term, or working memory. In memory-II, it is assumed that a certain memory pattern, $\overline{\mathcal{B}(t)}$, is formed which should resonate with input signals and further consolidate the memory. This memory is coined as an "Engram pattern" and will produce neural code, the Engram. The Engram is the basic neural code playing important roles in the neural circuit for cognitive signal processing.

*E-mail: zhcho36@gmail.com

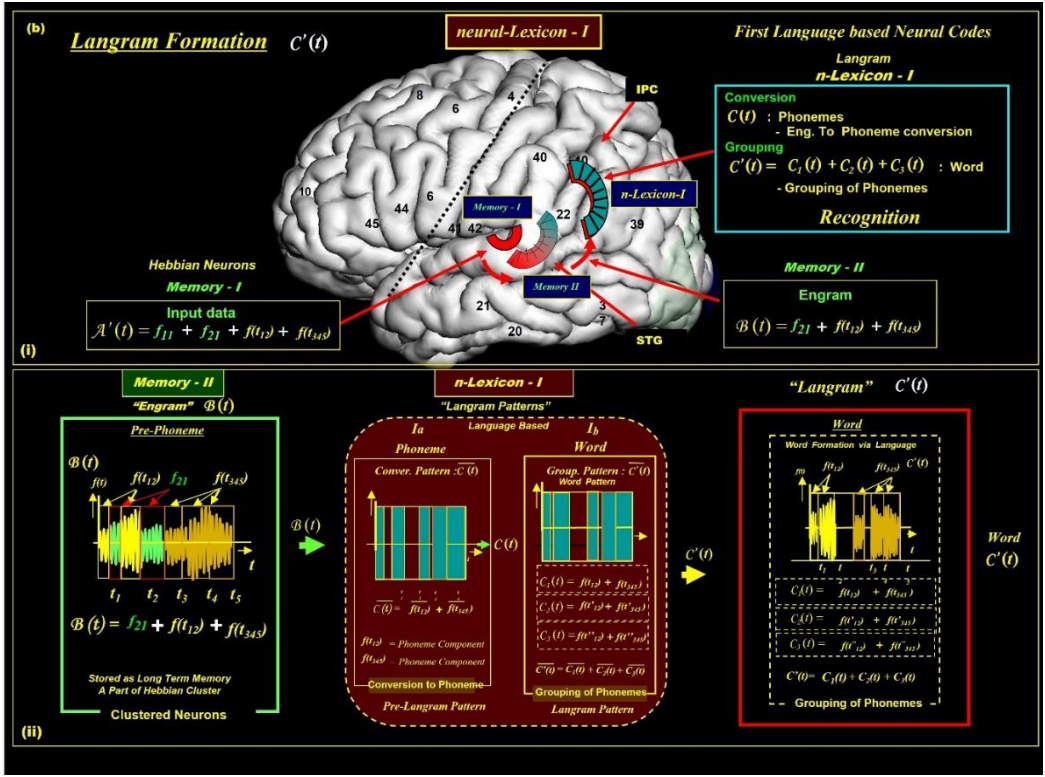

**(b).** The formation of the Engram is followed by the neural–Lexicons, which are the memory units developed via repeated language exposure and assumed to be located in higher cortical areas. The first neural Lexicon (n-Lexicon-I) is assumed to be located in the tertiary area in sensory sites, such as the inferior parietal cortex (IPC), the sensory input integrating area where the supramarginal and angular gyri are located. For this neural Lexicon, two major functions are envisaged; one for the absorption of the Engram for the conversion of the neural signal (Engram) into language-based neural code, the phonemes, and subsequent summing or grouping of the phonemes to form a Langram, word, or morpheme. These processes are illustrated at the bottom, and the resulting word or Langram is shown on the right. An interesting feature of the Langram is that it is the first neural code based on language used throughout the neural circuit leading to cognition and thinking.

Legends; $\overline{\mathcal{B}(t)}$: Engram pattern, a memory pattern for Pre-Phoneme. $\mathcal{B}(t)$: Engram (Pre-Phoneme). $\overline{\mathcal{C}(t)}, \overline{\mathcal{C}'(t)}$: Memory patterns for the Pre-Langram (Morpheme) and Langram (Word). $\mathcal{C}(t)$: Langram or word. $f(t_{12})$ and $f(t_{345})$: Components of a Phoneme. STG: Superior Temporal Gyrus. IPC: Inferior Parietal Cortex.

Subsequently, n-Lexicon-II involves another grouping process, as shown in Fig. 4. The result of the grouping of the words in n-Lexicon-II leads to a sentence or cognition, which we termed as "Cognogram" as shown in Fig. 4, which can be noted as,

$\mathcal{D}(t) = \mathcal{C}'_1(t) + \mathcal{C}'_2(t) + \mathcal{C}'_3(t)$: Phrase or Sentence (Group of words)...{Eq. 6} Cognogram

where $\mathcal{C}'_1(t)$, $\mathcal{C}'_2(t)$ and $\mathcal{C}'_3(t)$ are the words.

It should be noted that the n-Lexicon-II is a memory unit again with a pattern formed or developed via language. Incoming Langrams are grouped via neuronal resonance with a pattern formed at the n-Lexicon-II. We have drawn a group of fibers on top of the brain surface, which are the connecting pathways to the frontal areas from the sensory integration area, the inferior parietal cortex (IPC) to the prefrontal cortex (PFC), which we have obtained with 7T MRI fiber tractography (Cho[1] and see supplement III. Tractography). This shows how the sensory integration centers (IPC-supramarginal and angular gyri) are connected to the high cortical areas (PFC and orbitofrontal cortex [OFC]) in the frontal cortices, which are believed to be the central cognitive center[16–18].

**Approach to decision and thinking via neural Lexicons – n-Lexicon-III and beyond**

Further processing of the Cognogram in the n-Lexicon-III, located in the orbito-frontal cortex (OFC), constitutes the selection process. OFC is known to be closely related to the limbic system and, therefore, emotion. The limbic centers, such as the ACC (BA32, BA24), are believed to comprise old

*E-mail: zhcho36@gmail.com

memory units that act as a big memory bank. As shown in Fig. 5(a), n-Lexicon-III functions as a selection processor, with a selection pattern developed via language ($\overline{E(t)}$), once again, via neuronal resonance absorption of the incoming Cognogram. This results in a "decision" signal or "Decigram," which can be noted, for example, as,

$E(t) = C'_3(t)$: selected sentence, therefore, decision…{Eq.7} Decigram

This decision signal can lead to action or movement in the premotor and motor areas. However, another option would be the formation of the feedback loop by sending the Decigram signal, $E'(t)$, back to the n-Lexicon-II; this will form a circulating signal, the Circulogram, as shown in Fig. 5(b). The Circulogram at a certain time, that is $t = t_n$, $E''(t)$, leads to thinking, provided there are no more input sensory signals ($\mathcal{B}(t) = 0$), that is, for example,

$\boldsymbol{\Psi(t)} = E'(t)|_{\mathcal{B}(t)=0, t=t_n} = E''(t) = \boldsymbol{C'_1(t)}$: reselected Sentence, the "Thinking"…{Eq.8} Circulogram output at $t = t_n$

where $E'(t)$ and $E''(t)$ are the circulating or feedback signal and the output of the circulating signal, respectively.

In short, the neural signal flow for "cognition and thinking" can be summarized as follows: Sensogram – Engram – Langram (Pre-Langram and Langram) – Cognogram – Decigram – Circulogram. The Circulogram output at $t = t_n$, without the input signal to the Cognogram or *n-Lexicon-II*, leads to thinking, $\Psi(t)$.

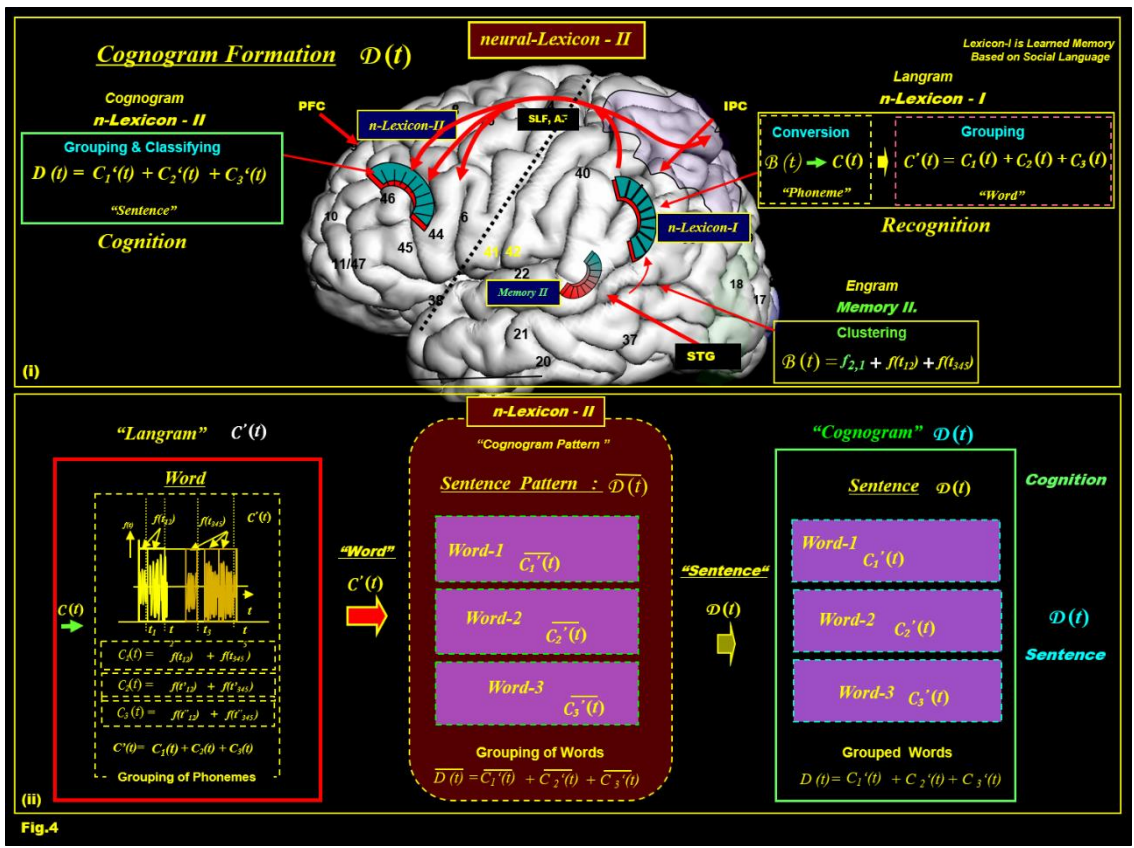

**Figure. 4**. After the Langram in the sensory integrating center (at the n-Lexicon-I) follows the transfer of the signals to the n-Lexicon-II in the cognitive center of the motor site, the Prefrontal Cortex (PFC). In the n-Lexicon-II, where the sentence pattern is developed by language and responds to the input Langrams or words, a sentence is formed by summing or grouping. The sentence is also known for "Integration or Classification," which leads to cognition processes involving such as comprehension. The process of cognition represents the major neural process in the brain and forms the bases for decisions and thinking.

Legends; $\overline{\mathcal{D}(t)}$: A memory pattern of a Sentence. $\mathcal{D}(t)$: Sentence or Cognogram. IPC: Inferior Parietal Cortex. STG: Superior Temporal Gyrus. PFC: Prefrontal Cortex.

*E-mail: zhcho36@gmail.com

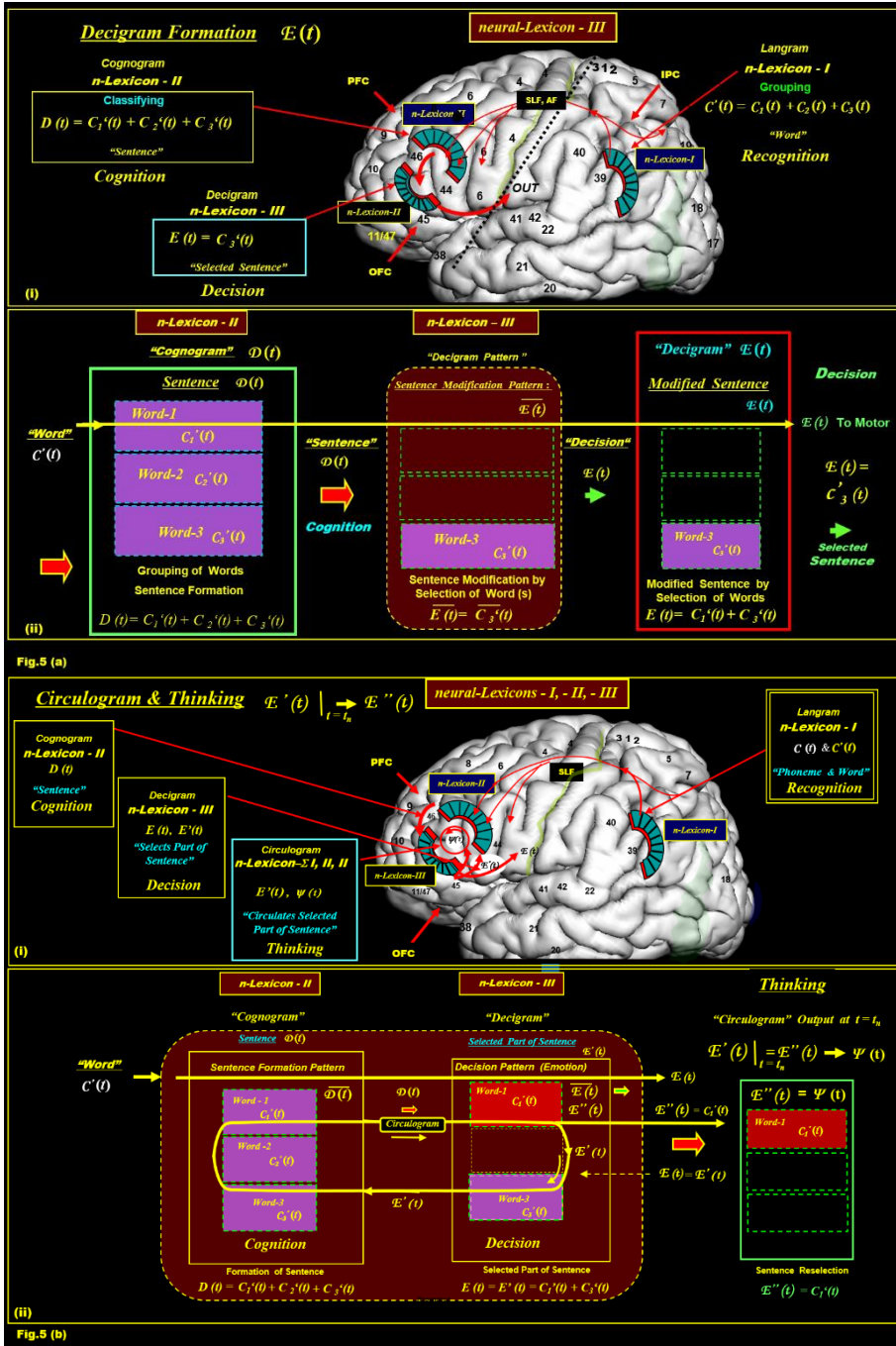

**Figure 5. (a).** In the n-Lexicon-III, we observe a selection or modification process by selecting some words from the sentence, which one can assume to be a decision process; that is, the input sentence is modified by the Decigram pattern, $\overline{E(t)}$, with which the incoming signal, Cognogram or sentence, is modified by partial selection of the content, resulting in a decision. The decision signal or Decigram will often be sent to the motor areas for the final action or movement. However, this could lead to the circulation or feedback loop and forms a Circulogram, with which thinking can be derived.

**(b).** At the last stage, the Decision signal can be fed back to n-Lexicon-II to form a feedback loop, a Circulogram. The Circulogram will lead to thinking. The output of the circulating signal at a certain time $t = t_n$, $E''(t)$, is assumed to be thinking, which we have also noted as $\Psi(t)$. Note C'$_1$(t), which is the second selection after the feedback and circulation.

Legends; $\overline{E(t)}$ : A memory pattern of Decigram. $E(t)$: Decigram or Decision signal. OFC: Orbitofrontal Cortex. STG: Superior Temporal Gyrus. IPC : Inferior Parietal Cortex. PFC: Prefrontal Cortex. $E'(t)$: A Circulogram. $E''(t)$ or $\Psi(t)$: Circulogram output, at $t = t_n$, the "Thinking." OFC : Orbitofrontal Cortex. PFC: Prefrontal Cortex. Note also that $E(t) = E'(t)$.

## 3. RESULTS

In conjunction with the model, whole cognitive neural signal processing is summarized and shown in **Fig. 6**, using the neural memories or patterns and proposed neural Lexicons. Starting from the Engram developed from the sensory input with the Engram pattern $\overline{\mathcal{B}(t)}$ at the left-most portion of the figure, we can observe the Engram $\mathcal{B}(t)$ entering into the first part of the neural Lexicon, *n-Lexicon-I*, where the phoneme pattern $\overline{\mathcal{C}(t)}$ is, and producing the first pre-language-based neural code, the pre-Langram or phoneme. Subsequently, a word is formed by grouping of phonemes $\mathcal{C}'(t)$. In the following stage for n-Lexicon-II, grouping of words form the sentence or phrase, which we call the Cognogram. We assumed its function to be cognition or comprehension. Sentence comprehension or cognition in linguistics is often known as classification as well.

*E-mail: zhcho36@gmail.com

Subsequent signal handling involves the processing of the output of cognition, the Cognogram, $\mathcal{D}(t)$. The output of cognition or the Cognogram is directed to *n-Lexicon-III*, where selection processes are performed, resulting in the decision signal or Decigram generation via a selection process. For each conversion, grouping, or selection process, memory patterns in the *n-Lexicons,* as well as neuronal resonance absorption or excitation[3,4] take place.

The next steps after the output of the decision process are diverse; for example, one directly leads to action and movement while the other can lead to feedback, which is send back to the cognition circuit and forms a feedback loop. The other could lead to the formation of the output from the feedback loop to create a feedback loop output at a specific time, for example, $t = t_n$, the thinking, $\Psi(t)$.

The first action or movement coordination is the most common behavioral result via the premotor and motor cortices. The second is the feedback loop formation, and it is another necessary process that can lead to a higher cognitive process, the Circulogram. The third process, which is the formation of the output of the feedback loop or Circulogram output, can be expected at a particular time. For this, which is the Circulogram output at a specific time $t = t_n$, we propose as the thinking process, which is one of the highest cognitive processes that can be expected in the brain. We have assigned this process as $E''(t)$, as shown in **Fig. 6.**

In short, the entire neural signal process can be summarized as; First, the input signal to the first n-Lexicon in the system is assumed as an Engram. At the *n-Lexicon-I*, the Engram is converted to language-based neural code, the pre-Langram or phoneme, and subsequently grouped into the Langram or word. A phoneme is a phonologic term with several components, such as $f(t_{12})$ and $f(t_{345})$, as shown in Fig. 2(b), where the Engram consists of several phonologic components, as well as a noise component, $f_{21}$. The Engram is, therefore, termed a pre-phoneme.

After the Langram or word formation at *n-Lexicon-I (I_b)*, further signal processing is followed at the *n-Lexicon-II* where another grouping is performed. This produces the sentence, a group of morphemes or words also called Langrams. We termed this sentence formation the cognition or Cognogram. Cognition is undoubtedly one of the most significant processes through which we comprehend, classify, and, thereby, make decisions based on the selection process of the classified data, as shown in the last part of the n-Lexicon series, the *n-Lexicon-III*.

A decision signal or the Decigram has three directions; the premotor and motor cortices for action and movement; return to the *n-Lexicon-II,* the PFC, for the feedback loop formation; and the output formation from the feedback loop, Circulogram output, or thinking, respectively. These three outputs, which are action and movement, feedback loop formation, and output of feedback loop, represent significant human cognitive activity, including the immediate response, halting or deciphering, and conclusion after thinking. Note that these processes involve language-based neural Lexicons, suggesting the importance of language in human cognitive processing. Also note that the final product, which is thinking, is without sensory input, especially to the *n-Lexicon-II,* the PFC. An old quotation is that "thinking" is "…. coherent happening in the brain without sensory input…."

As noticed, the modeling and analysis of the neural circuit we have defined are based on a few new recent developments in neuroscience, namely: the Engram theory and the Hebbian neuronal process by Semon and Hebb and Tonegawa[2,5,6]; neural resonance theory proposed by Hutcheon and Matsumoto[3,4]; modern neural network theory of deep learning[14]; mental Lexicon[19] concept; recent development of high-resolution fiber tractography[1,20]; and the newly proposed neural Lexicon hypothesis involving Language proposed by us.

## 4. DISCUSSION

The present modeling and results obtained are based on modern neuroscience concepts and experimental data (tractography) and neural evidences, such as the Engram and neuronal resonance. However, the overall system-wise modeling combining the deep learning concept and neuroscience-based neural Lexicon hypothesis are new, and we expect the model to be a valuable tool for future language and cognitive neuroscience studies, as well as for the burgeoning AI research.

Cognition, in terms of our human neural circuitry, mainly based on human neurobiological bases, has been discussed by several papers[11,17,21–33]. However, further studies are required because the findings are not conclusive.

We are now better equipped theoretically and experimentally to test the "neural Lexicon" hypothesis, and, therefore, the neural network model

*E-mail: zhcho36@gmail.com*

of "cognition and language" or "cognition and thinking," because of several ingenious studies, such as those on the neuronal bases of the Engram and neuronal resonance, as well as the new high resolution in-vivo human fiber tractography or human brain fiber connectivity data[1,3,4,6,28,30,34].

Cognition is also possible in non-human primates, but probably on a much-reduced scale because their lexicons are not formed as richly due to the absence of human-like languages. In non-human primates, cognitive capacity or activity is limited due to poorly formed neural Lexicons because of the lack of languages, and their brain structures are not as developed for the formation of the Circulogram like in humans. Another way of saying this is that animal cognition and Decigram output are connected mainly to the action and movement channel rather than the feedback loop for the Circulogram or thinking. It has been generally assumed that thinking is performed in the prefrontal areas, which means that the Circulogram is in the *n-Lexicon-II* and *–III*. Thus, the richness of the neural Lexicons developed via languages and the formation of the Circulogram are two unique human properties that lead to cognition and thinking; therefore, the main difference between animals and humans appears to be related to the richness of their neural Lexicons, and, directly and indirectly, via languages.

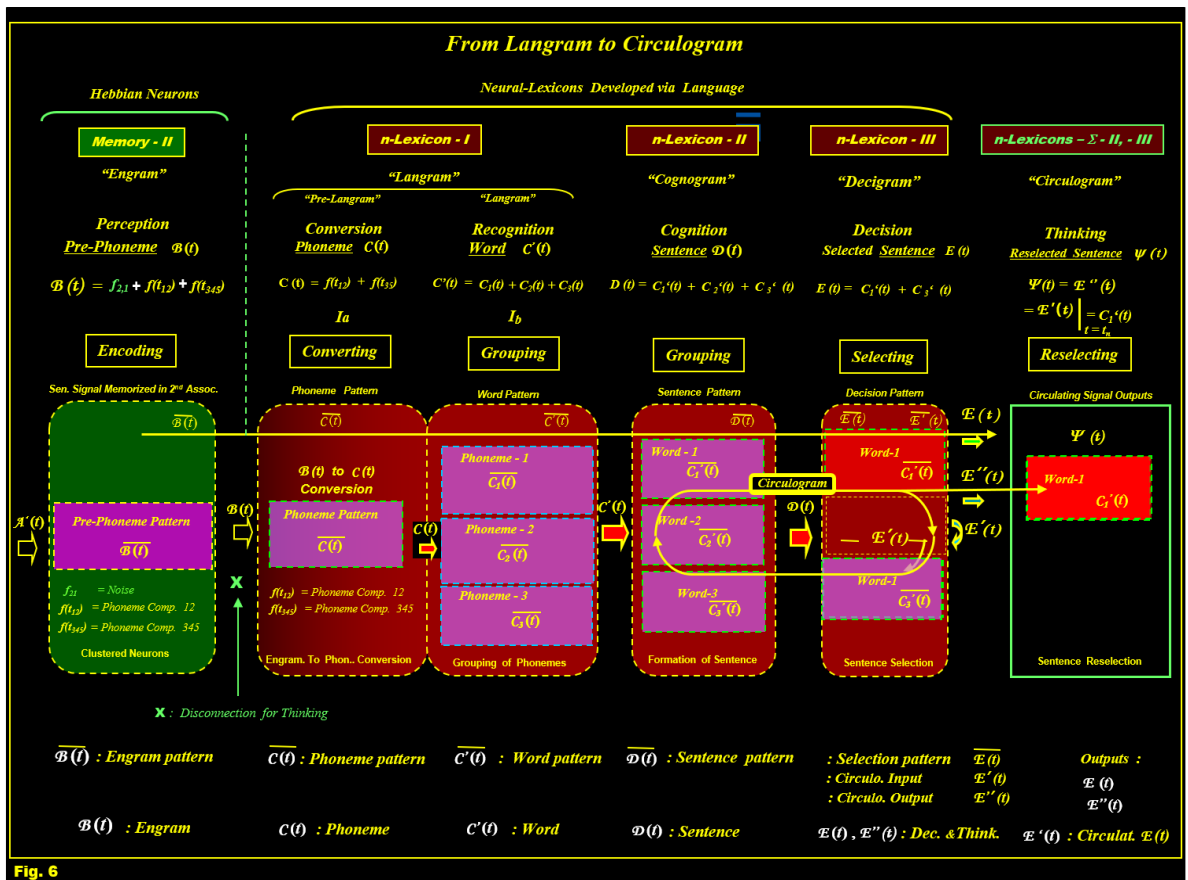

**Figure 6.** Summary of the five subdivisions of the neural signal processes shown in Fig. 3-5. The formation of the Circulogram through the neural Lexicons (n-Lexicon - II, and - III) will provide an output signal Decigram or the Circulogram, $E'(t)$. This output of Circulogram, at $t = t_n$, or $E''(t)$, $\Psi(t)$, is termed as "*Thinking*." "*Thinking*" is considered as "…coherent happening in the brain without sensory input." We have, therefore, assumed that the input to the system, the Engram, therefore, Langram is no longer presents when the Circulogram is activated. "*Thinking*" is a result of the circulation of the Decigram over the n-Lexicon-II and -III, which involves only the higher brain areas: the motor tertiary areas or the prefrontal cortex (PFC)and the orbitofrontal cortex (OFC). The limbic area is believed to be involved in decision-making.

Legends; $\overline{\mathcal{B}(t)}$: Engram pattern. $\overline{\mathcal{C}(t)}, \overline{\mathcal{C}'(t)}, \overline{\mathcal{D}(t)}, \overline{E(t)}, \overline{E'(t)}$: Language-based memory patterns of the neural lexicons; the Pre-Langram, Langram, Cognogram, Decigram, and Circulogram patterns, respectively. $E'(t)$: A Circulogram. $E''(t)$ or $\Psi(t)$: Circulogram output, "Thinking."

*E-mail: zhcho36@gmail.com

## 5. CONFLICT OF INTEREST

The authors declare that the research was conducted without any commercial or financial relationships construed as a potential conflict of interest.

## 6. AUTHOR CONTRIBUTIONS

ZH Cho formulated the hypothesis and drafted the manuscript. SH Paek, TY Cho, YB Kim, HJ Jeong and HG Lee contributed to the various aspects of neuro-anatomical and functional studies, the overall research support of the program, and suggested the integration of the AI and deep learning concept to the study.

## 7. FUNDING

This work was supported by the Brain Research Program of the National Research Foundation of Korea (NRF), funded by the Ministry of Science and ICT (2017M3C7A1049026).

## 8. ACKNOWLEDGEMENTS

The authors would like to thank their colleagues and friends who have provided valuable suggestions and advice and participated in discussions about Engram, Langram, lexicons, and languages.

## 9. DATA AVAILABILITY STATEMENTS

The raw data supporting the conclusions of this article will be made available by the authors without undue reservation.

*E-mail: zhcho36@gmail.com

# *Supplementary Material*

**1    Supplementary Figures and Tables**

**1.1   Supplementary Figures**

*I . System Flow Diagram (Figs. 1, 2)*

*II . Neuronal Resonance (Figs. 3, 4, 5)*

*III. Tractography (Figs. 6, 7, 8, 9, 10)*

*IV. Engram (Figs. 11, 12)*

*V . Deep Learning Model (Fig. 13)*

*VI . Conceptual Flow Diagram of Human Cognition and Thinking (Fig. 14)*

*VII. Table 1. Key Scientific Components, New Terminologies and Keywords*

**I.  System Flow Diagram (Details)**

*Fig. 1.  Over all signal flow diagram of the "Cognition".*

*Fig. 2.  Deep Learning like Neural-Lexicon (n-Lexicon)*

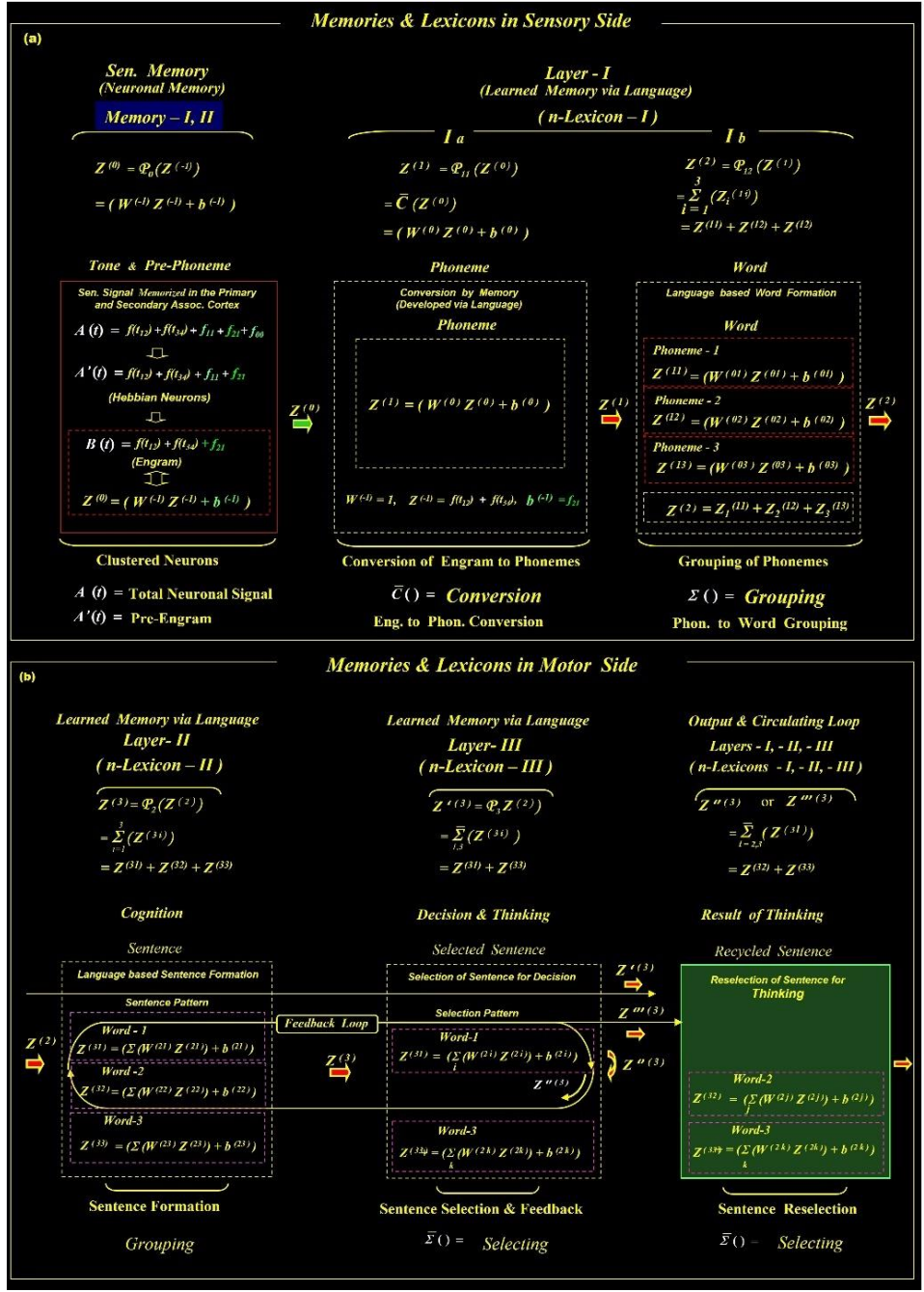

**Supplementary Figure 1.** Overall signal flow diagram of the "Cognition".

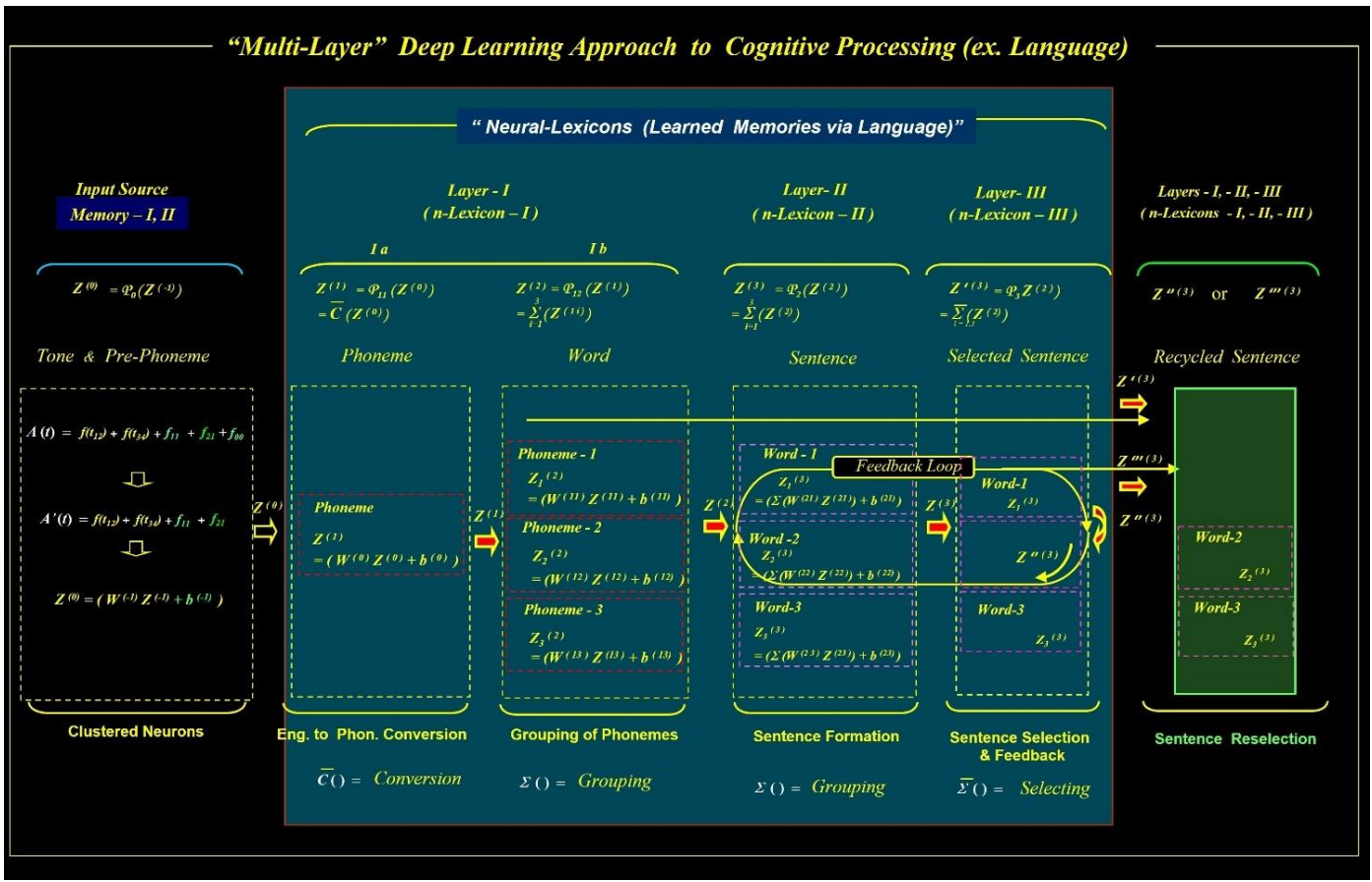

**Supplementary Figure 2.** Deep Learning like Neural-Lexicon (n-Lexicon).

## II. Neuronal Resonance

"Resonance" is Everywhere, Nature's Most Efficient Tool for Communication, Filtration, Classification or Rearrangement, and Selection.

"Resonance" helps to make "Neuronal Communication"

"Resonance" helps communication between the Sensogram, Engram, Langram, Cognogram, Decigram, & Circulogram

*Fig. 3. Neuronal Resonance and Its Electrical Circuit Equivalent*

*Fig. 4. Y. Matsumoto-Makitono et al. Cell Reports, Jul 26, 16(4): 994-1004, 2016*

*Fig. 5. Neuronal Resonance and Neuronal Communication*

**References**

*B. Hutcheon and Y. Yarom. Trends in Neuroscience, TINS Vol. 23, No. 5, 216-222, 2000*

*Y. Matsumoto-Makitono et al. Cell Reports, Jul 26, 16(4): 994-1004, 2016*

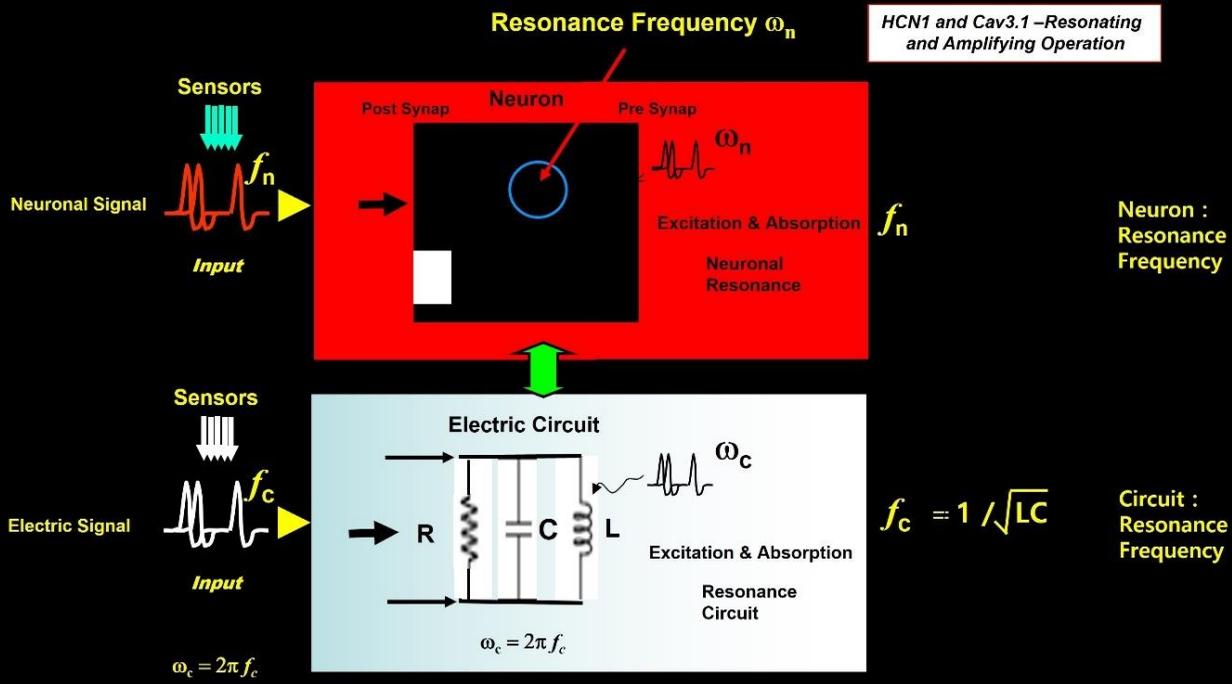

**Supplementary Figure 3.** Neuronal Resonance and Its Electrical Circuit Equivalent

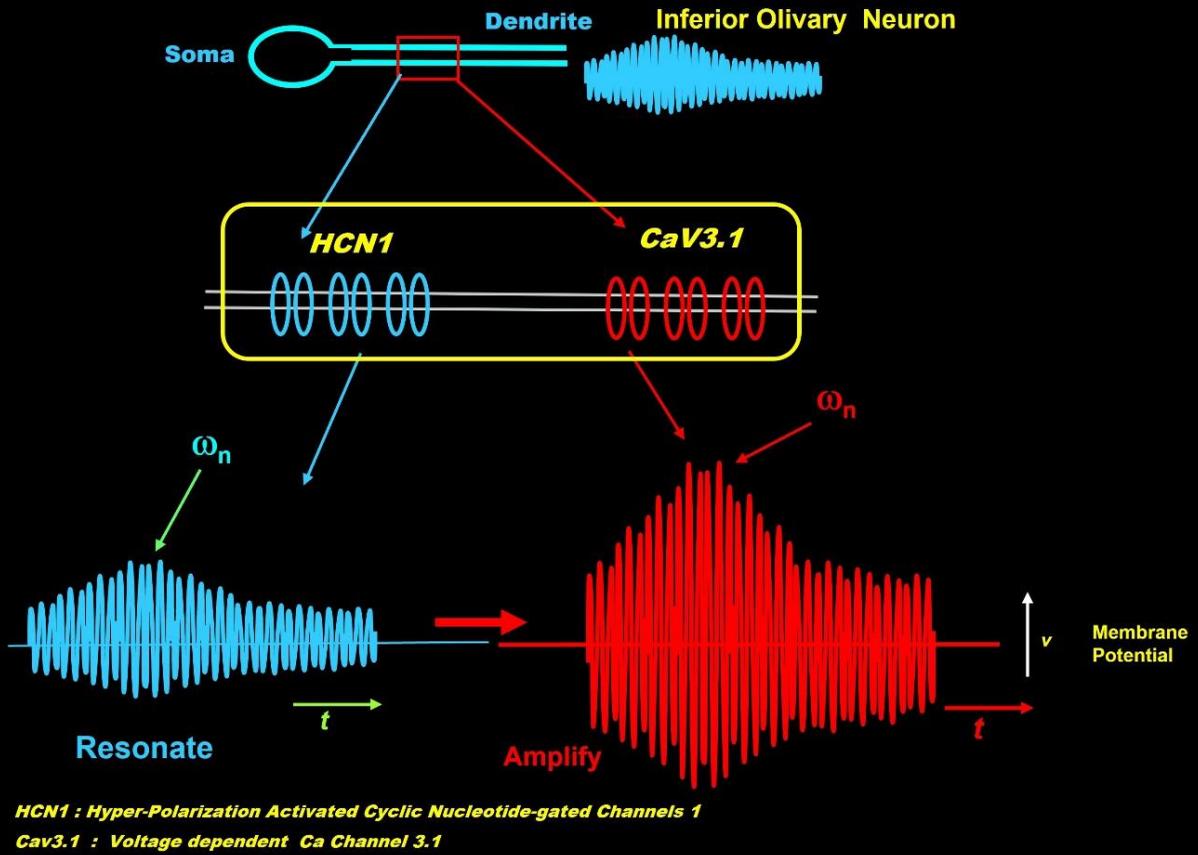

**Supplementary Figure 4.** Y. Matsumoto-Makitono et al. Cell Reports, Jul 26, 16(4): 994-1004, 2016

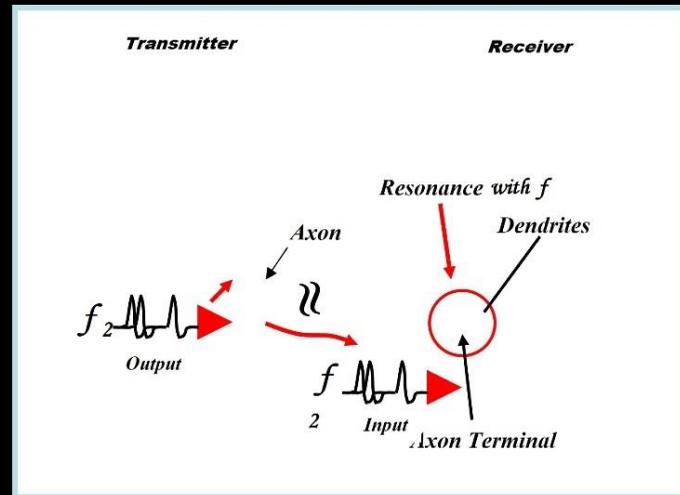

**Supplementary Figure 5.** Neuronal Resonance and Neuronal Communication

### III. Tractography

#### Advanced 7T MRI Diffusion Tensor Imaging

New Tractography with Ultra-High Field 7.0T MRI began to provide the "Connectivity Map" between the sensory areas to the motorareas thereby enable us to hypothesize quantitatively the "Functional Connectivity" map hitherto unable to do with available techniques, for example fMRI.

*Fig. 6. 7.0T MIR Diffusion Tensor Imaging data with post processing*

*Fig. 7. Selected Fiber Tracks of the 7T MRI Diffusion Tensor Imaging*

*Fig. 8. 10-Division Model tractography of the Left Hemisphere*

*Fig. 9. Fibers Superimposed on to the Brainnetome Atlas*

*Fig. 10. 10-Division Model compared with the previously published data*

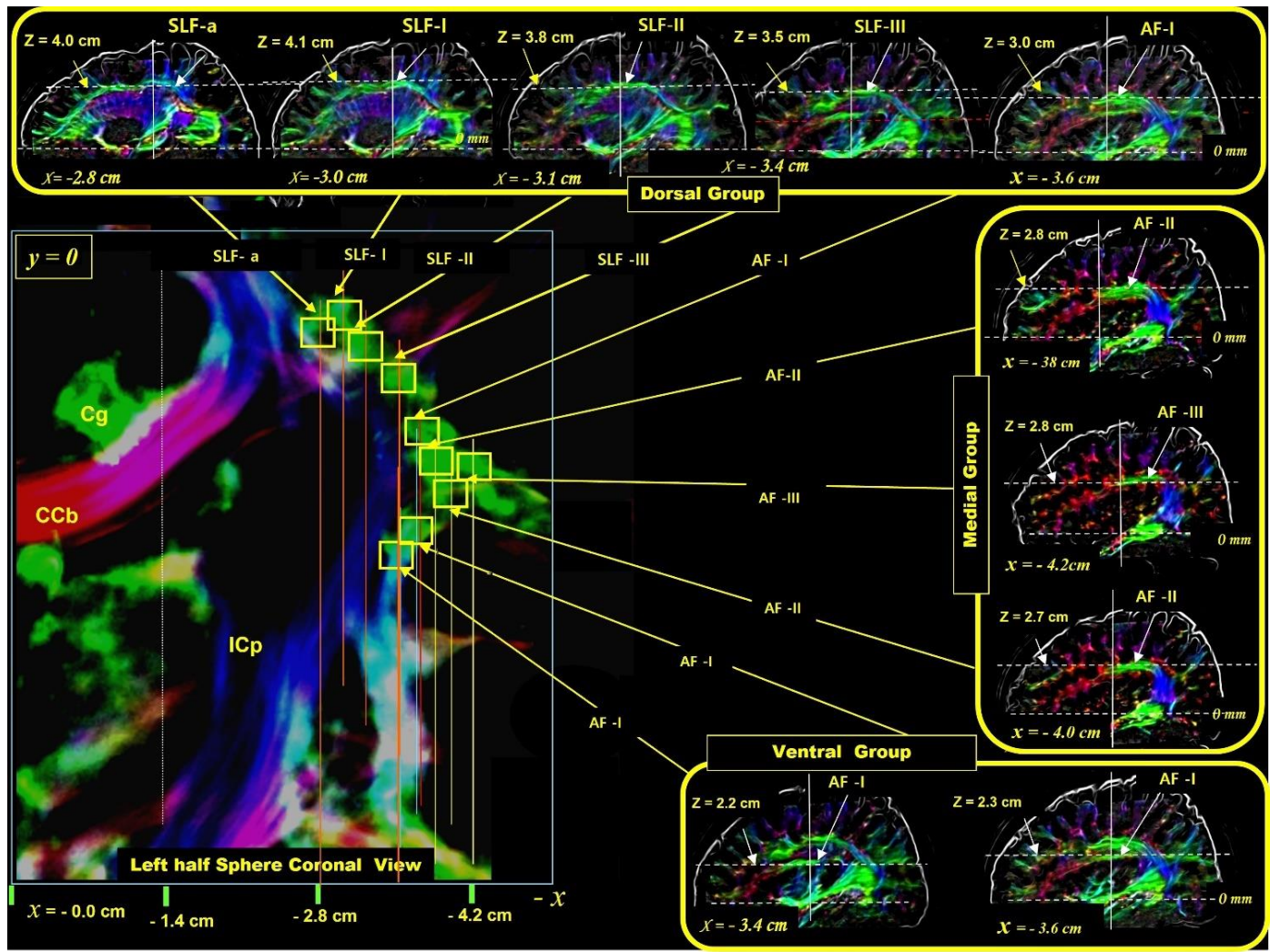

**Supplementary Figure 6.** 7.0T MIR Diffusion Tensor Imaging data with post processing

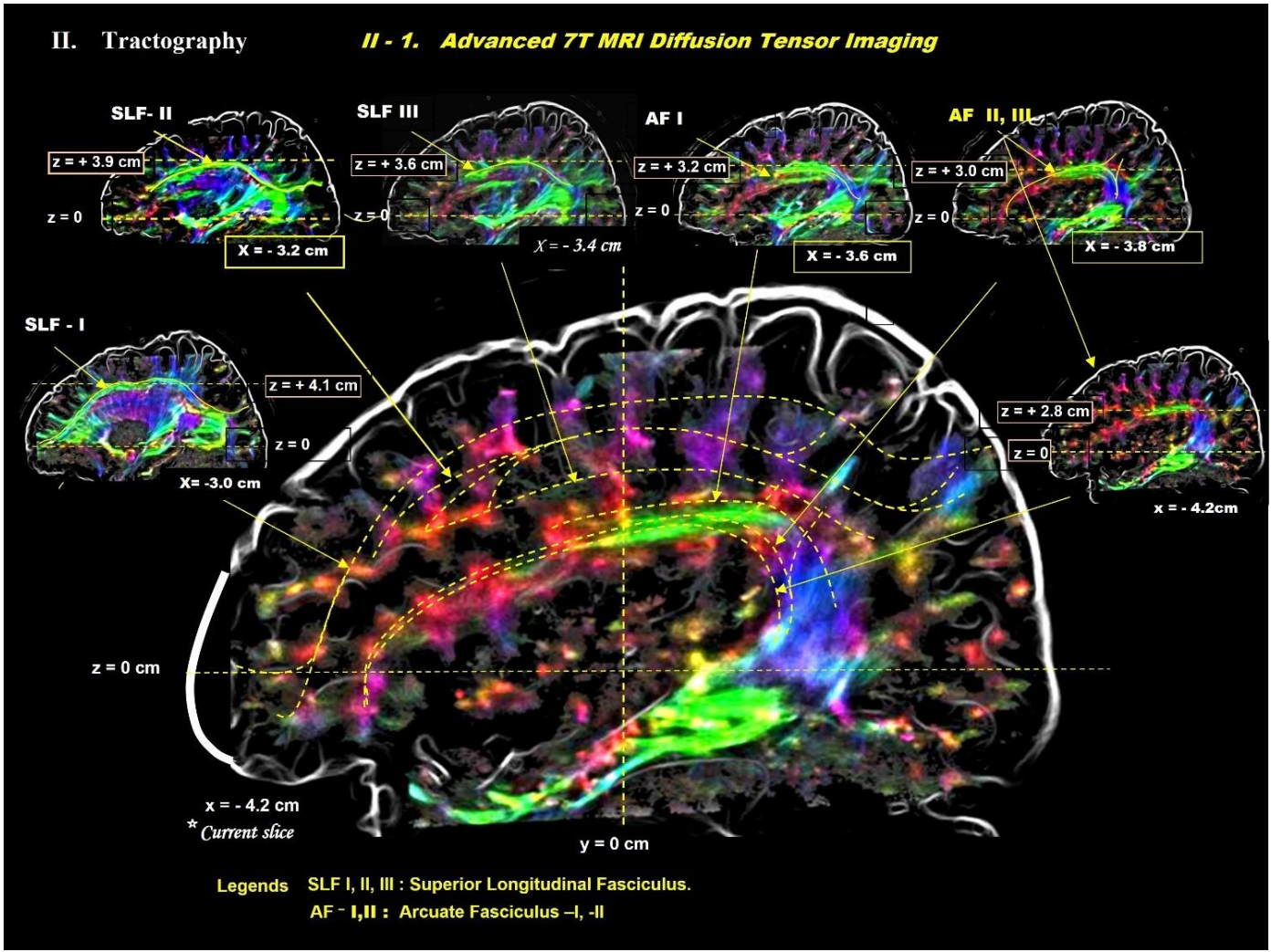

**Supplementary Figure 7.** Selected Fiber Tracks of the 7T MRI Diffusion Tensor Imaging

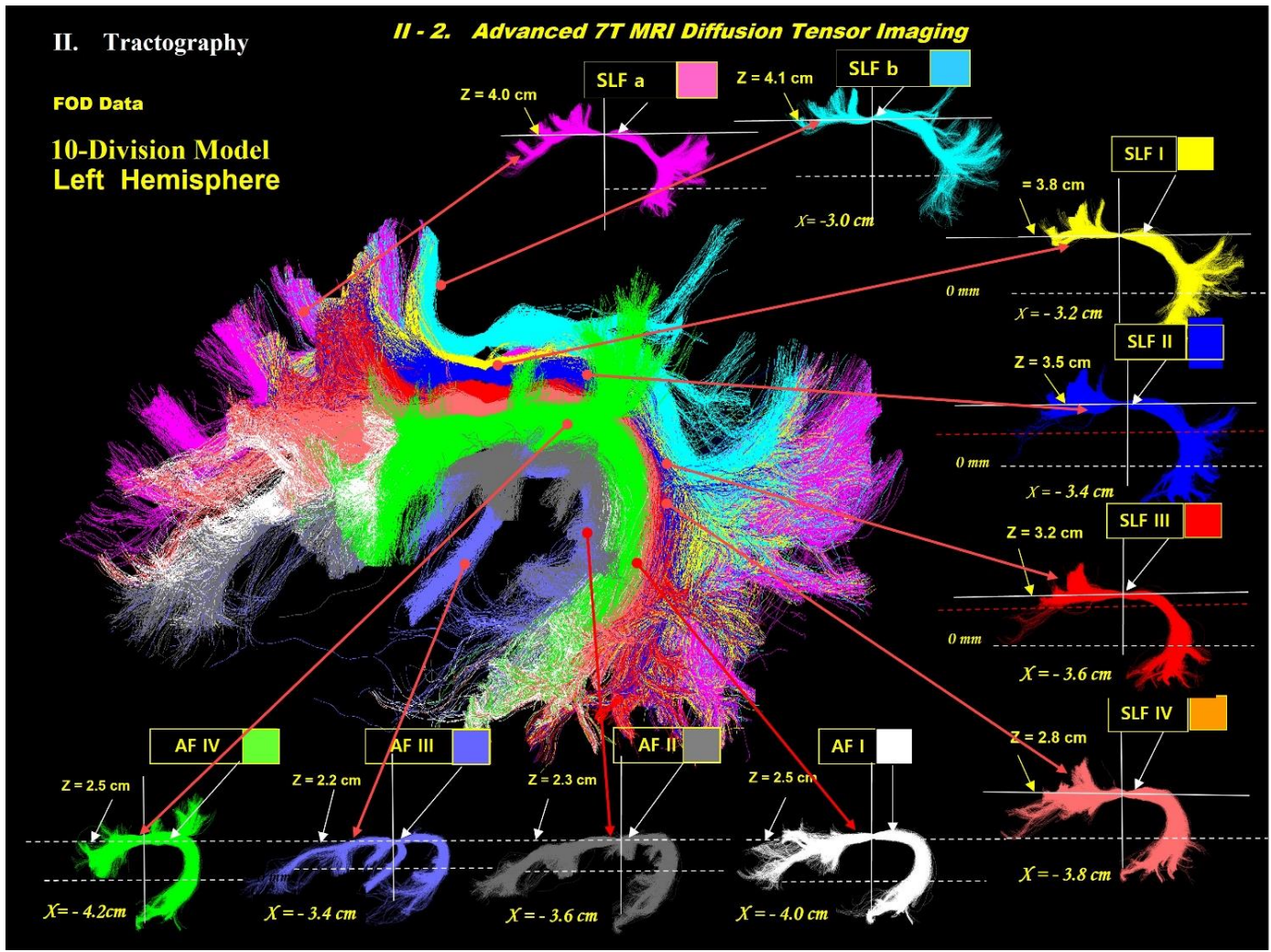

**Supplementary Figure 8.** 10-Division Model tractography of the Left Hemisphere

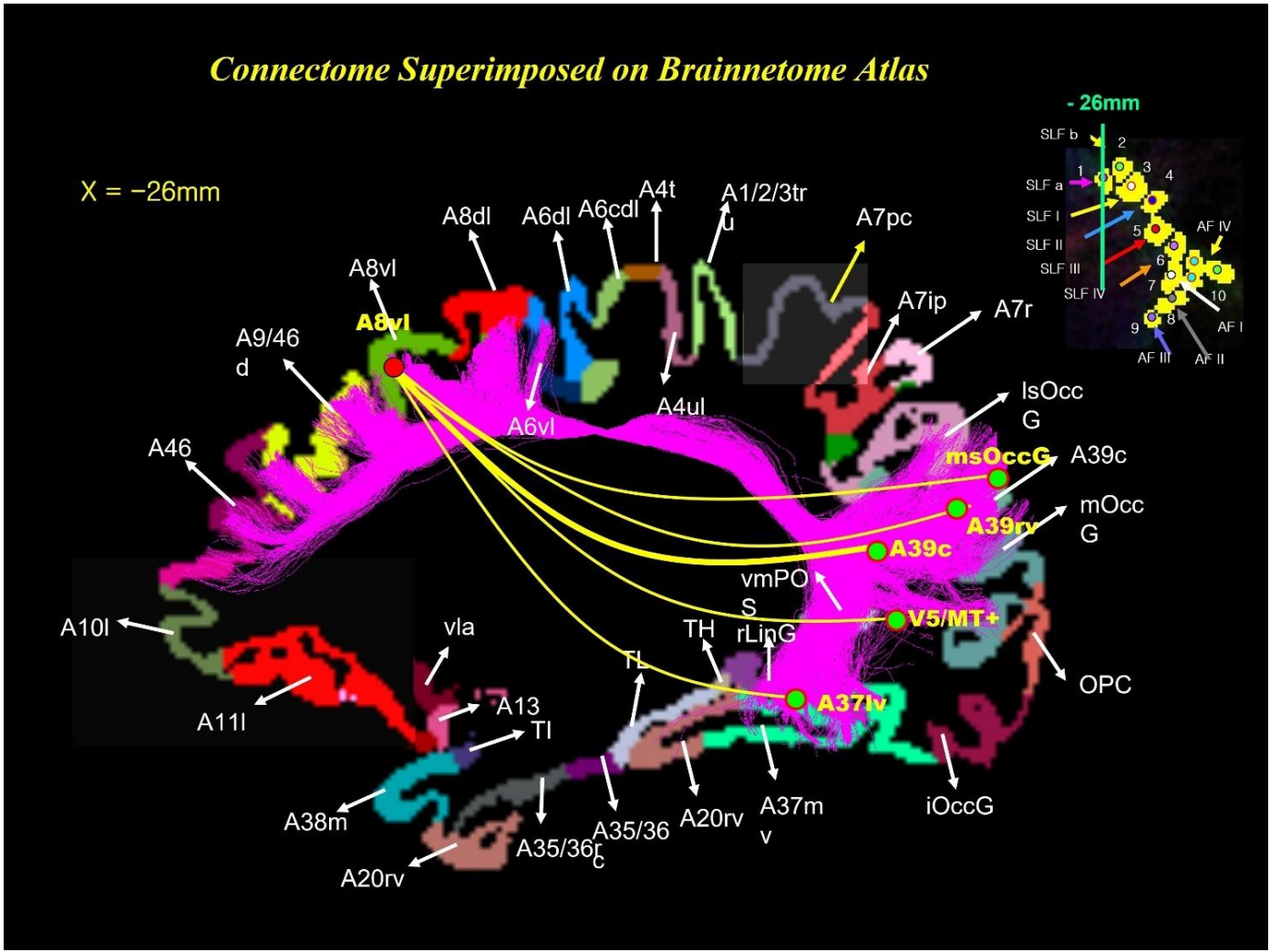

**Supplementary Figure 9.** Fibers Superimposed on to the Brainnetome Atlas

**Supplementary Figure 10.** 10-Division Model compared with the previously published data

## IV.   Engram

**Neural Code Associated with Intermediate to Long Term Memory at the Secondary Sensory Association Cortex or Hippocampus.**

*Fig. 11. llustration of an Engram*

*Fig. 12. Engram and Resonance Excitation*

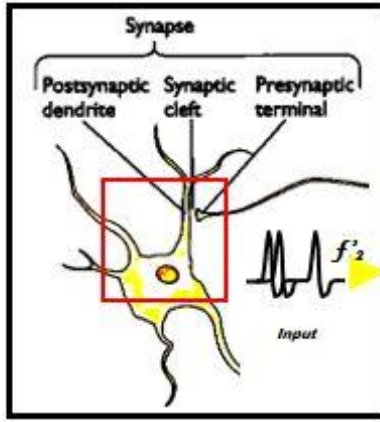
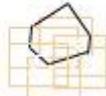
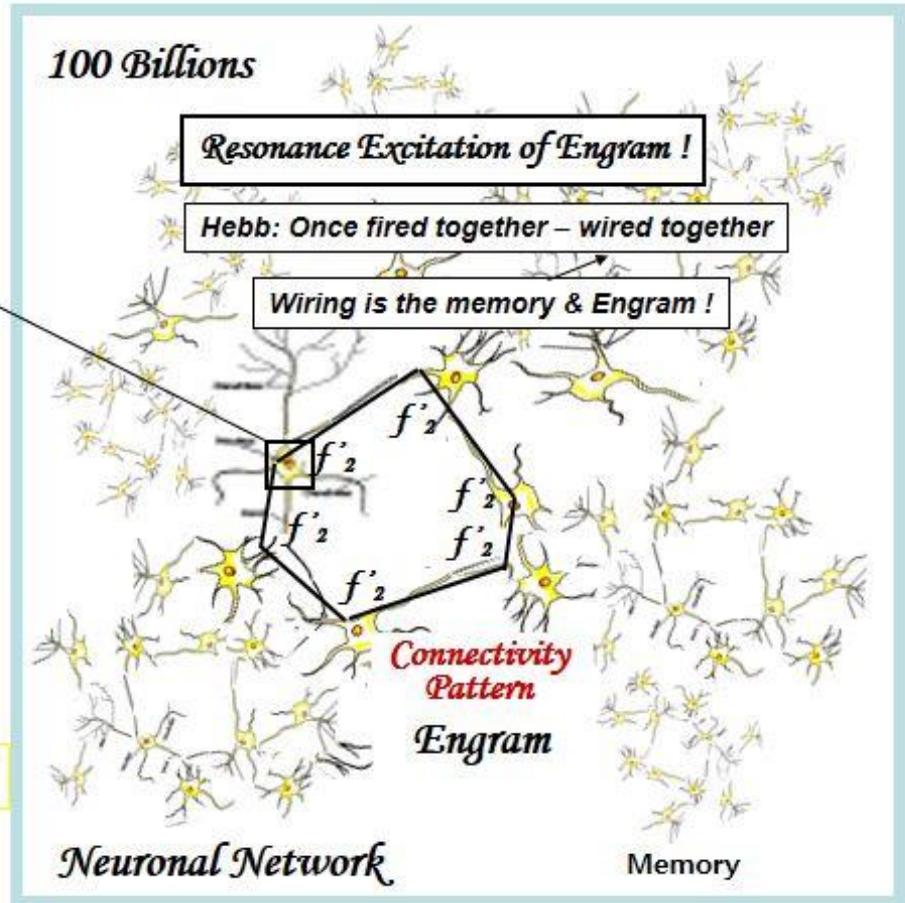

**Supplementary Figure 11.** Illustration of an Engram

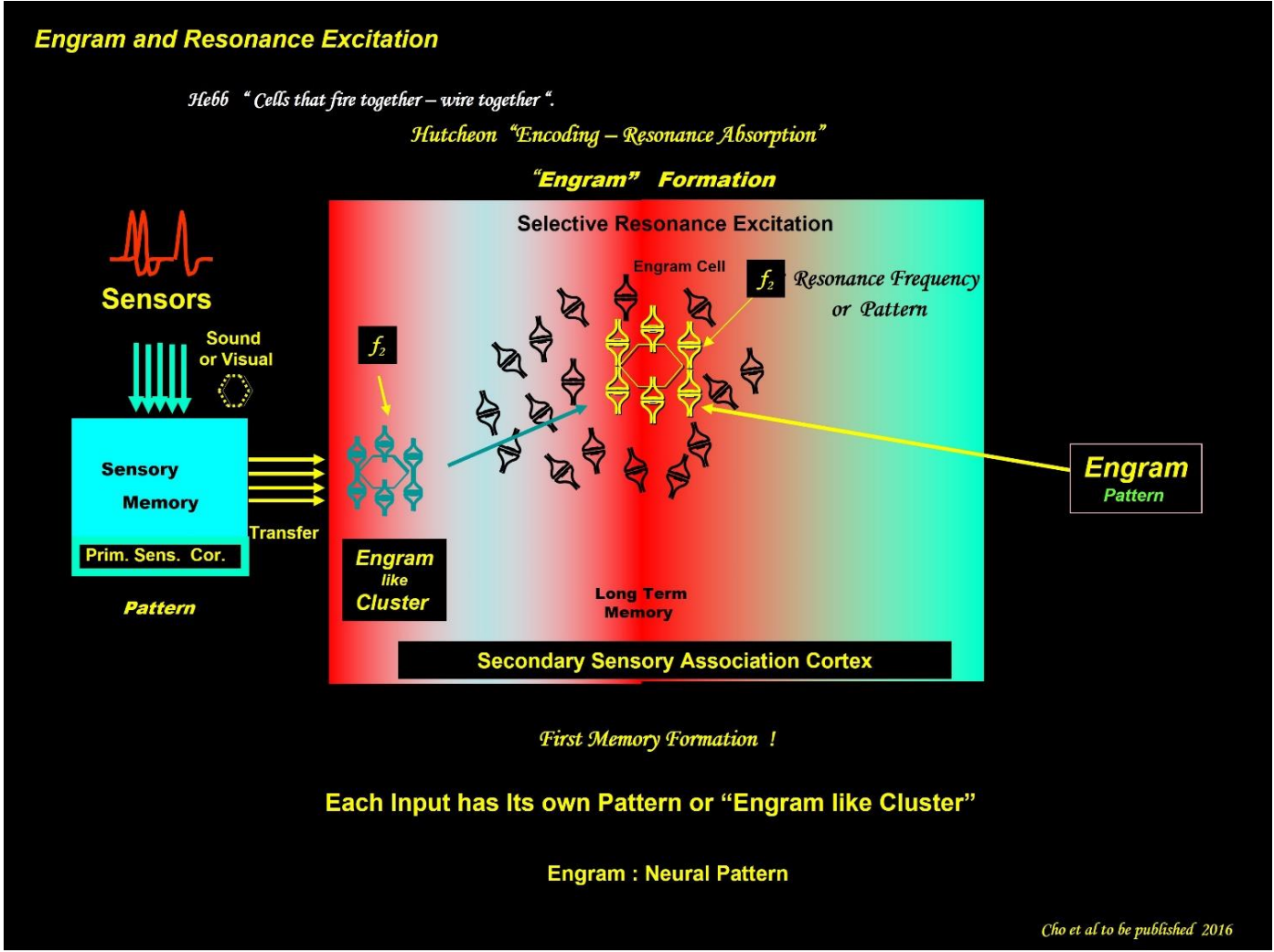

**Supplementary Figure 12.** Engram and Resonance Excitation

**V.  Deep Learning Model**

**Deep Learning has been one of the leading theoretical model for the Artificial Intelligence (AI) and neural computation.**

*Fig.13. Deep Learning Model*

**References**

Schmidhuber J. (2015). Deep Learning in Neural Networks : An Overview. Neural Netw. 61, 85-117.

**Supplementary Figure 13.** Deep Learning Model

## VI. Conceptual Flow Diagram of Human Cognition and Thinking

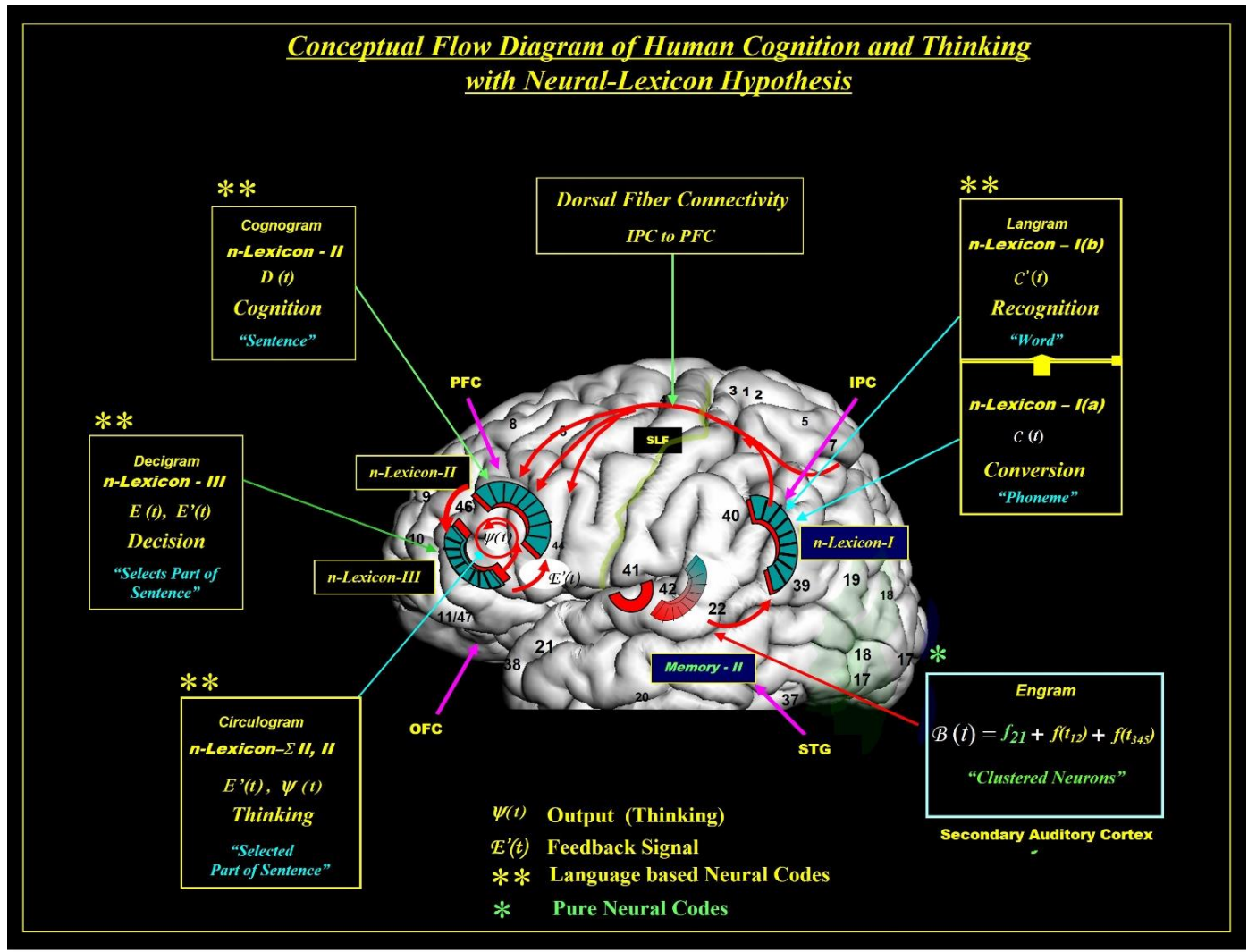

**Supplementary Figure 14.** Conceptual Flow Diagram of Human Cognition and Thinking with Neural-Lexicon Hypothesis

## 1.2 Supplementary Tables

**Supplementary Table 1.** Key Scientific Components, New Terminologies and Keywords

**Key Scientific Components**

(a). Neural-Lexicons (see Ref. Miozzo M. 2000)

(b). Neuronal Resonance (see Ref. Hutcheon B. 2000, Mutsumoto-Makidono Y. 2016)

(c). Engram, Hebbian Clustering, Hebbian Neurons (see Ref. Semon R. 1904, Hebb D.O. 1949, Tonegawa S. 2015)

(d). New High Resolution Fiber Tractography (see Ref. Cho Z.H. 2015)

(e). Deep Learning Model for n-Lexicons (see Ref. Schmidhuber J. 2015)

**New Terminologies :**

(a). Sensogram (Sensory-Engram).

    Engram : Pre-Engram, Engram (Pre-Phoneme).

        Langram (Language-Engram), Word : Pre-Langram (Phoneme).

            Cognogram (Cognition-Engram), Sentence.

                Decigram (Decision-Engram), Decision.

                    Circulogram (Circulation-Engram). Thinking.

(b) Neural-Lexicons (n-Lexicons)

    n-Lexicon-I, n-Lexicin-II, n-Lexicon-III

**Keywords :**

Neuronal - Primary or Secondary, Neural- High Level or Tertiary, Neural Memory, Neuronal Pattern, Language based Neuronal Pattern, Neural (pure) Code (Engram), Language based Neural Code (Pre-Langram, Langram …), Conversion of Neuronal Code to Language based Neural Code (ex. Engram to Pre-Langram), Learned Memory via Language (n-Lexicon), Neural Memory via Language (n-Lexicon), Neural Memory Developed by Learning via Language (n-Lexicon), Neural-Lexicon (n-Lexicon); Grouping of Language based Neural Codes (ex. Pre-Langrams to Langrams or Langrams to Cognogram), Selection of Language based Neural Codes (Selection of a Certain Number of Langrams in Cognogram and Decigram). Feedback Loop, Circulogram